\documentclass[fleqn,usenatbib]{mnras}

\usepackage{newtxtext,newtxmath}
\usepackage[T1]{fontenc}

\DeclareRobustCommand{\VAN}[3]{#2}
\let\VANthebibliography\thebibliography
\def\thebibliography{\DeclareRobustCommand{\VAN}[3]{##3}\VANthebibliography}

\usepackage{graphicx}	% Including figure files
\usepackage{amsmath}	% Advanced maths commands
\usepackage{orcidlink}

\title[Photometry and model of 2025~FA$_{22}$]{Photometry and physical characterization of near-Earth asteroid 2025~FA$_{22}$ from one apparition}
\author[J. Tian et al.]{
Jun Tian,$^{1}$\orcidlink{0000-0001-9341-4007}
Bin Li,$^{1}$\orcidlink{0000-0001-9327-0920}
Y. J. Liu,$^{1}$
Z. J. Xu,$^{1}$
R. Y. Zhai,$^{1,2}$
H. B. Zhao,$^{1}$ \thanks{E-mail: meteorzh@pmo.ac.cn}\orcidlink{0000-0001-9059-5623}
Anton Pomazan,$^{3}$
Fan Li,$^{1}$ 
\newauthor A.M. Abdelaziz,$^{4}$
Y.D. Ping,$^{1}$
Wei Liu,$^{1}$
Y.D. Mao,$^{3}$
Chen Zhang, $^{1}$
Jian Chen, $^{1,2}$
Ahmed. Shokry,$^{4}$
\newauthor and Mohamed Ismail$^{4}$
\\
$^{1}$Purple Mountain Observatory, Chinese Academy of Sciences, Nanjing 210023\\
$^{2}$School of Astronomy and Space Sciences, University of Science and Technology of China, Hefei 230026, China \\
$^{3}$Shanghai Astronomical Observatory, Chinese Academy of Sciences, Shanghai, China\\
$^{4}$National Research Institute of Astronomy and Geophysics (NRIAG), Helwan, Cairo 11421, Egypt \\
}
\date{Accepted XXX. Received YYY; in original form ZZZ}

\pubyear{\the\year{}}

\begin{document}
\label{firstpage}
\pagerange{\pageref{firstpage}--\pageref{lastpage}}
\maketitle

\begin{abstract}

We present comprehensive photometric characterisation of 2025~FA$_{22}$, a Potentially Hazardous Asteroid (PHA) discovered on 29 March 2025 and observed during the seventh International Asteroid Warning Network (IAWN) coordinated campaign. The asteroid's close approach at 2 lunar distances on 18 September 2025 provided an opportunity for rapid physical characterisation in a simulated virtual impactor scenario. Photometric observations were conducted from 17 September to 1 October 2025, during which 2025~FA$_{22}$ traversed a $150^{\circ}$ arc and spanned solar phase angles from $\sim20^{\circ}$ to $\sim70^{\circ}$. This geometry enabled detailed physical characterisation, including determination of the spin axis and shape. Convex inversion yields a sidereal rotation period of $P_{\mathrm{sid}} = 13.07366 \pm 0.00076$~h and a spin axis at ecliptic coordinates $({\lambda}, {\beta}) = (246^{\circ} \pm 9^{\circ}, 60^{\circ} \pm 9^{\circ})$. The absolute magnitude was derived as $H_{\mathrm{V}} = 21.39^{+0.07}_{-0.08}$~mag ($G_1 = 0.8228$, $G_2 = 0.0194$), with colour indices $B-V = 0.71 \pm 0.05$~mag, $V-R = 0.39 \pm 0.03$~mag, and $R-I = 0.39 \pm 0.04$~mag, consistent with X-complex classification in the Bus--DeMeo taxonomy. Assuming a geometric albedo of $p_v = 0.15^{+0.05}_{-0.04}$, representative of the moderate-albedo X-complex asteroids, we estimate $D_{\mathrm{eff}} = 
181^{+31}_{-25}$~m, consistent with ``China Compound Eye'' radar dimensions of ${\sim}100 \times 320$~m ($D_{\mathrm{eff}} \sim 186$~m), which also reveal a contact-binary morphology consistent with the axis ratios $a{:}b{:}c \sim 2.68{:}1.96{:}1.00$ derived from our photometric shape model. The bilobate morphology is consistent with YORP-driven spin-up and deformation during a previous YORP cycle, with the current slow rotation possibly explained by internal reconfiguration or tidal braking during close Earth encounters.

\end{abstract}


%%%%%%%%%%%%%%%%%%%%%%%%%%%%%%%%%%%%%%%%%%%%%%%%%%

%%%%%%%%%%%%%%%%% BODY OF PAPER %%%%%%%%%%%%%%%%%%

\section{Introduction}

Near-Earth asteroids (NEAs) are a critical population of Solar System bodies with orbits in close proximity to Earth. The physical characterization of NEAs is therefore essential both for small-body science and planetary defence \citep{Fatka2022}. Photometric observations provide lightcurves, phase curves, and multi-band colours for NEAs. These data constrain rotation periods, spin-axis orientations, shapes \citep{Kaasa2004,durech2015}, albedos \citep{Arcoverde2023}, and taxonomic types \citep{Hromakina2023,pereira2025}. Additionally, such observations enable the assessment of the Yarkovsky--O'Keefe--Radzievskii--Paddack (YORP) effects that influence asteroid rotational evolution \citep{Wlodarczyk2026,2003Natur}.

Comprehensive characterization of asteroid physical properties typically requires multi-apparition photometric observations spanning years or decades \citep{kaasalainen2001a, Durech2020}. Each apparition provides observations from different viewing geometries, which are crucial for constraining shape models and pole orientations through lightcurve inversion techniques \citep{kaasalainen2001a, durech2010}. However, single-apparition observations face inherent geometric limitations: the restricted range of solar phase angles and aspect angles during a single observing window often provides insufficient constraints for unique determination of rotation state and shape, particularly for newly discovered objects. For planetary defence, the ability to characterize potentially hazardous asteroids (PHAs) from single-apparition observations is critically important for rapid threat assessment and response planning \citep{devog2025}.When a new object poses a potential threat, we need timely constraints on rotation period, shape, spectral type, and size to inform impact analyses and mitigation strategies. Multi-year campaigns are incompatible with such scenarios, making intensive single-apparition observations particularly valuable.

The YORP effect is particularly relevant for contact binaries—bilobed NEAs that provide strong constraints on small-body structure and rotational evolution. Radar observations suggest that approximately $15\text{--}30~\%$ of NEAs with diameters $>200~\mathrm{m}$ are contact binaries \citep{2015aste.book..165B,Virkki2022}. YORP-driven spin-up is commonly invoked as an efficient pathway to produce such configurations \citep{walsh2008,campo2020}; however, the existence of slowly rotating systems ($P > 10~\mathrm{h}$) indicates that additional formation channels may be at play, including low-velocity mergers \citep{Johansen2015} and BYORP-related evolution \citep{Cuk2005,McMahon2010}. Whether contact binaries form predominantly from two separate objects or represent the natural evolution of a single body remains unclear. To date, only 16 near-Earth contact binaries have been characterized with detailed physical parameters and reliable shape models \citep{Cannon2025}. A larger sample is required to refine evolutionary models and distinguish between formation mechanisms.

The PHA 2025 $\mathrm{FA}_{22}$ is an Apollo-group asteroid discovered by Pan-STARRS 2 at Haleakala on 29 March 2025. The discovery was reported in the MPEC 2025-G circular\footnote{\url{https://www.minorplanetcenter.net/mpec/K25/K25G41.html}}. It will make a close approach to Earth on 18 September 2025 at approximately 2 lunar distances ($\sim$768,000 km). The IAWN designated this encounter as its seventh coordinated campaign to test global observational capabilities in obtaining precise astrometric and physical characterization data for asteroid impact scenarios\footnote{\url{https://iawn.net/obscamp/2025FA22/}}. The campaign allows us to assess, under operationally realistic conditions, how well single-apparition photometry can constrain and characterise the rotational state and physical properties of a newly discovered PHA.

We conducted a multi-site ground-based photometric campaign with an extended temporal baseline to obtain high-precision light curves of 2025 $\mathrm{FA}_{22}$ before and during its close approach. This work contributes to the IAWN campaign by providing data for physical characterization. We present a comprehensive photometric characterization of 2025 $\mathrm{FA}_{22}$, with the main goal of constraining its spin, shape, and related physical properties. The paper is structured as follows: Section 2 describes the photometric observations of 2025 $\mathrm{FA}_{22}$ and the data reduction procedures; Section 3 presents lightcurve analysis and shape modeling results; Section 4 analyzes the taxonomic classification and the Magnitude-phase curve, from which the absolute magnitude, the G parameter, and other related physical properties are derived. An analysis of the rotational state of 2025 $\mathrm{FA}_{22}$ and a discussion of possible formation mechanisms for its contact-binary configuration are presented in Section 5; and a summary of our key results and future work is presented in Section 6.

\section{Observations and data reduction}
\subsection{Observations and Instruments}

We organized an observing campaign for 2025 $\mathrm{FA}_{22}$, focusing on multi-colour and lightcurve observations. By coordinating observation schedules across seven stations, we established an observing network spanning from Purple Mountain Observatory, XuYi Station (MPC code: D29) in the east to ShAO Chile Station, El Sauce (MPC code: X08) in the west. The spread in longitude enabled near-continuous coverage of the asteroid's apparent motion. Details of all observing systems are provided in Table~\ref{tab:1}.

\begin{table*}
\caption{Details of observatories, telescopes and detectors used in observations.}
\begin{tabular}{lllllll} % four columns, alignment for each
\hline
Observatory & MPC code & Telescope & Focal & Pixel scale & FOV & Filter$^{*}$\\
\hline
Purple Mountain Obs., Jiama'erdeng Station, China & N54 & 0.7 m RC & 7.0 m & 0.44$^{\prime\prime}$/pixel& $15^{\prime} \times 15^{\prime}$ & C/u/g/r/i/z  \\
& & 0.6 m & 1.02 m & 0.76$^{\prime\prime}$/pixel& $181^{\prime} \times 136^{\prime}$ & C/r \\
& & 0.4 m RC & 0.88 m & 0.506$^{\prime\prime}$/pixel& $81^{\prime} \times 54.2^{\prime}$ & C/r/i \\
Purple Mountain Obs., Yaoan Station, China & O49 & YAHPT & 8.0 m & 0.385$^{\prime\prime}$/pixel& $13.1^{\prime} \times 13.1^{\prime}$ & C/B/V/R/I \\
Purple Mountain Obs., Lenghu Station, China & O17 & 0.8 m RC & 5.6 m & 0.14$^{\prime\prime}$/pixel& $22.2^{\prime} \times 14.8^{\prime}$ & C/g/r/i \\
Shanghai Astron. Obs., El Sauce, Chile & X08 & 0.5 m & 2.93 m & 0.63$^{\prime\prime}$/pixel & $43.2^{\prime} \times 43.4^{\prime}$ & C \\
Gaoyazi Obs., Xinjiang, China & -- & 0.5 m & 1.63 m & 0.481$^{\prime\prime}$/pixel & $76.6^{\prime} \times 51.1^{\prime}$ & C \\
Kottamia Obs., Helwan, Egypt & 088 & 1.88 m RC & 11.54 m & 0.25$^{\prime\prime}$/pixel& $8^{\prime}\times8^{\prime}$ & U/B/V/R/I/u/g/r/i/z \\
Purple Mountain Obs., Xuyi Station, China & D29 & CNEOST & 1.8 m & 1.03$^{\prime\prime}$/pixel& $181.2^{\prime} \times 181.2^{\prime}$ & B/V/R/I/g/r/i \\
\hline
\end{tabular}
\label{tab:1}
\raggedright
\footnotesize
\begin{minipage}{\textwidth}
$^{*}$ Filters U/B/V/R/I are from the Johnson-Cousins system, u/g/r/i/z from the SDSS system, and C denotes a clear filter (unfiltered).
\end{minipage}
\end{table*}

\begin{figure}
	\includegraphics[width=0.95\columnwidth]{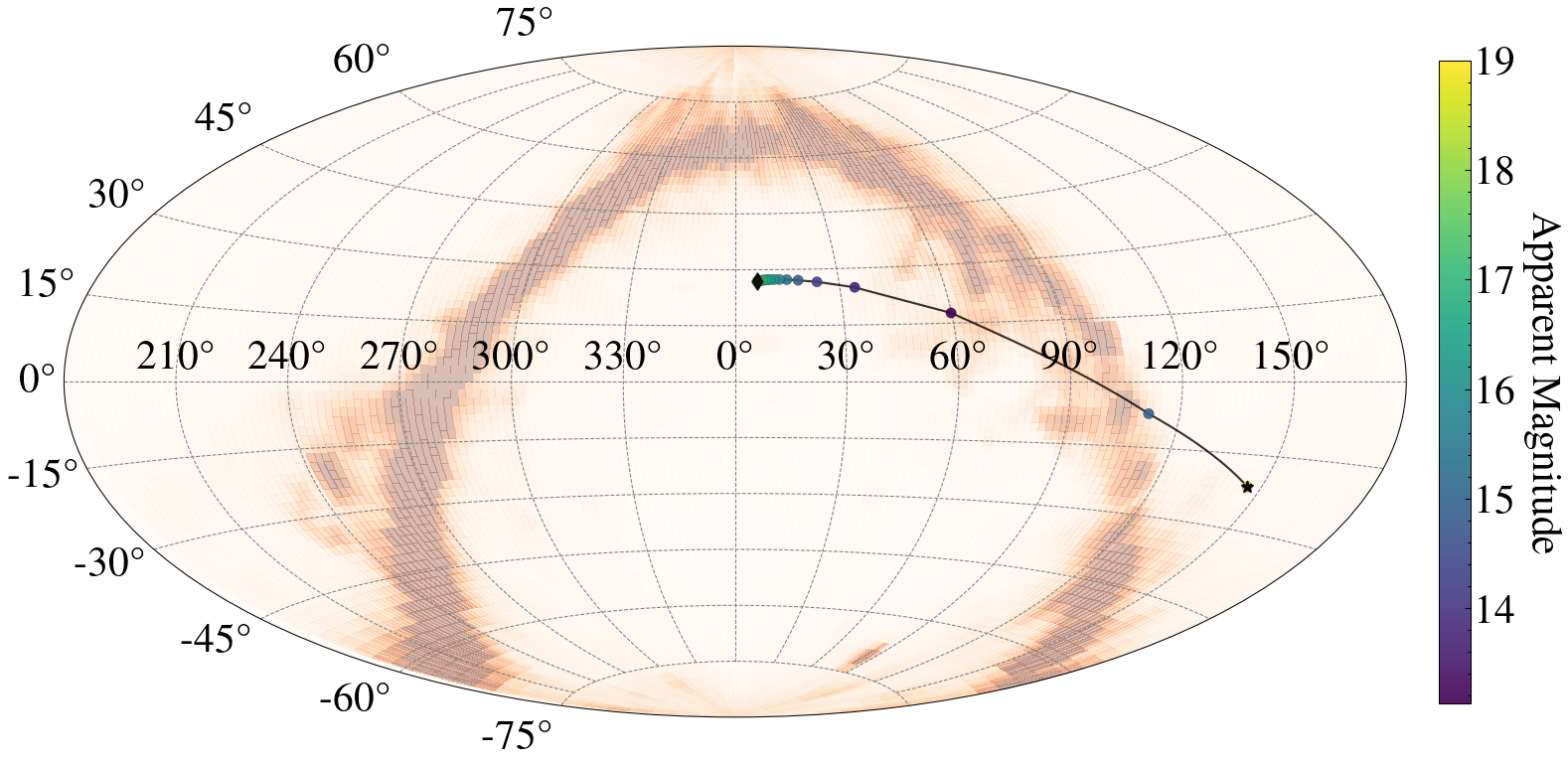}
    \caption{Sky-plane trajectory made by 2025 $\mathrm{FA}_{22}$ during our observing campaign, plotted in equatorial (RA, DEC) coordinates.Observations span from 17 September to 4 October 2025 (UTC), covering an arc of approximately $\sim$150$^{\circ}$ on the sky, with colours indicating the apparent magnitude variation along the trajectory. The colour scale represents the apparent magnitude. The green star indicates the start of the observing sequence, and the red diamond marks the end. Note that observations from 17 September and 4 October were excluded from the lightcurve inversion analysis due to low signal-to-noise ratio.}
    \label{fig:1}
\end{figure}

The Minor Planet Center ephemeris\footnote{\url{https://minorplanetcenter.net/db\_search/show\_object?utf8=\%E2\%9C\%93&object\_id=2025+FA22}} indicated a peak apparent motion of 150~arcsec~min$^{-1}$ for 2025 $\mathrm{FA}_{22}$ during close approach, requiring short exposures with careful tracking. The object was observable at V~$\lesssim$~18~mag from 17 September to 4 October 2025, spanning $\sim$150$^{\circ}$ with phase angles 22$^{\circ}$--69$^{\circ}$ (Fig.~\ref{fig:1}). This observing geometry allowed us to determine the rotation period and colour indices. It also provided an opportunity to derive the magnitude-phase relation, spin axis orientation, and shape model.

Observations used non-sidereal tracking where possible. Under sidereal tracking, exposures were limited to $T_{\rm exp} \leq 2^{\prime\prime}/v$ ($v$ in arcsec~s$^{-1}$) to avoid trailing. Targets were observed above 30$^{\circ}$ elevation at $V \lesssim 18$~mag. Lightcurve data were obtained in R, r, or clear filters; B/V/R/I and g/r/i/z multi-colour photometry for colour-index determination. Observations were suspended during poor weather or strong moonlight. XuYi Station (MPC code: D29) was weathered out. Kottamia Observatory (MPC code: 088) data from 19 September were discarded due to trailing.

Photometric observations of 2025 FA$_{22}$ were conducted over 25 nights between 17 September and 1 October 2025 (UTC) using multiple telescopes. The observations including astrometry, multi-colour photometry, and lightcurve monitoring. The multi-colour observations followed the sequence B-V-B-V-R-V-R-I-R-I or g-r-g-r-i-r-i-z-i-z, designed to minimize rotational effects on colour-index determination \citep{mofan2025}. The astrometric data were submitted to the IAWN Astrometry Working Group and are not discussed further. 

\subsection{Data Reduction}

Raw frames were processed using standard reduction techniques, including bias correction and normalization with flat-field images (and for the dark current, if necessary). The image timestamps were adjusted to represent mid-exposure times. All photometric data were reduced using a Python-based automated pipeline adapted from \citet{mom2017}. Aperture photometry was performed on the frames. Despite the fast sky motion of the asteroid (up to $\sim$ 115~arcsec~min$^{-1}$), the point spread function (PSF) remained nearly circular even with 4 s exposures, showing no significant trailing. 

We extracted sources from raw frames using {\sc SExtractor} \citep{SEX}, then cross-matched them with the Gaia DR3 catalog \citep{GAIA3} through {\sc SCAMP} \citep{scmap}or Astrometry.net\footnote{\url{https://nova.astrometry.net/}} for astrometric calibration. Target positions were computed from JPL Horizons ephemeris\footnote{\url{https://ssd.jpl.nasa.gov/horizons/}} and matched to the source catalog within 2~arcsec to identify each asteroid. For photometric calibration, clear-filter observations were calibrated against Gaia G-band magnitudes, while standard-filter observations used their respective photometric systems. Calibration stars were selected with colour indices within 0.2~mag of solar colours \citep{mom2017}, using at least three stars per frame. Aperture photometry was performed with identical apertures for the asteroid and reference stars.

After reduction, we keep only measurements with signal-to-noise ratios (SNR) above 50. The 0.7-m data were affected by a filter-system problem that renders the multi-band measurements unreliable. This problem does not affect the r band; we therefore discard the intended g,r,i and z colour measurements and retain only the r-band time-series photometry for further analysis. The 0.4-m observations on 2025 September 18 show larger scatter due to the asteroid's fast sky motion and changes in wind speed during the night. We remove only data points with photometric uncertainties > 0.1~mag and keep the rest.

Photometric data were binned to 1-min intervals. For each epoch, we computed light-time corrections and observing geometry using a Python pipeline interfacing with the JPL Horizons ephemeris service. This procedure provided the observer-centred ecliptic longitude and latitude, heliocentric and geocentric distances, solar phase angle, as well as the the asteroid-centric ecliptic Cartesian coordinates of the Sun and observer (in au). These parameters define the viewing and illumination geometry required for lightcurve inversion. 

For convex inversion with {\sc DAMIT}\footnote{\url{https://damit.cuni.cz/projects/damit/}}, reduced magnitudes were converted to relative intensities and formatted according to the {\sc DAMIT} input specifications. After applying these corrections, a total of 22 independent lightcurves were obtained and used in the subsequent analysis. Table~\ref{tab:2} summarizes the final set of lightcurves together with the corresponding observing geometry. 

\begin{table*}
\centering
\caption{A log of optical photometry data sets of asteroid 2025 $\mathrm{FA}_{22}$ used in this study. Each line represents a single-night light curve identified by a numerical ID. The Universal Time (UT) ‘Date’ of the beginning of the night is given, as well as the heliocentric (r) and geocentric ($\Delta$) distances measured in au, the solar phase angle ($\alpha$), the observer-centred ecliptic longitude ($\lambda_{\mathrm{O}}$), the observer-centred ecliptic latitude ($\beta_{\mathrm{O}}$) at the mid-time of the observing night, the predicted average V magnitude from the Horizons ephemeris, and the MPC code of the observing station. The table also gives the asteroid movement on the sky (Motion), the total observing duration (Total), the exposure time (Exposure), and the filter used in the observations. }
\begin{tabular}{llcccccccccccc} % four columns, alignment for each
\hline
IDs&UT Date&r & $\Delta$ & $\alpha$&$\lambda_{\mathrm{O}}$ &$\beta_{\mathrm{O}}$&Motion&V&Total &Filter & Exposure$^{\dagger}$ &MPC code \\
&[yyyy-mm-dd]&[au]&[au]&[$^{\circ}$]&[$^{\circ}$]&[$^{\circ}$]&["/min]&[mag]&[h]&&[s]&\\
\hline
1&2025-09-18&1.0072&0.0067&68.11&64.64&-4.68&115.11&13.14&2.0&r&4&N54 \\
2&2025-09-19&1.0096&0.0086&54.40&51.38&4.23&70.78&13.27&1.6&C&2&X08\\
3&2025-09-20&1.0129&0.0118&44.09&40.84&10.58&38.66&13.68&1.5&R&6&088\\
4&2025-09-20&1.0146&0.0137&40.87&37.30&12.85&29.10&13.90&1.9&C&2&X08\\
5&2025-09-20&1.0164&0.0155&38.43&34.62&14.22&22.45&14.12&2.6&R&4&O49\\
6&2025-09-21&1.0216&0.0213&33.75&29.22&17.13&12.25&14.69&1.7&R&4&O49\\
7&2025-09-21&1.0223&0.0220&33.29&28.69&17.38&11.61&14.75&2.9&C&4&$*$\\
8&2025-09-21&1.0226&0.0224&33.11&28.45&17.52&11.51&14.78&8.2&r&10&N54\\
9&2025-09-22&1.0282&0.0287&30.46&25.40&19.03&7.27&15.25&8.2&r&10&N54\\
10&2025-09-23&1.0337&0.0347&28.71&23.49&19.92&5.14&15.63&8.3&r&15&N54\\
11&2025-09-24&1.0390&0.0407&27.38&22.16&20.51&3.89&15.94&8.7&r&30&N54\\
12&2025-09-24&1.0378&0.0393&27.68&22.47&20.36&3.78&15.88&0.4&B/V/R/I&15&O49\\
13&2025-09-25&1.0446&0.0468&26.26&21.12&20.94&3.04&16.23&8.7&r&40&N54\\
14&2025-09-25&1.0435&0.0456&26.47&21.33&20.83&2.80&16.17&0.8&r&40&O17\\
15&2025-09-26&1.0501&0.0529&25.31&20.31&21.24&2.45&16.48&8.5&r&50&N54\\
16&2025-09-27&1.0559&0.0593&24.44&19.63&21.48&2.03&16.71&8.6&r&60&N54\\
17&2025-09-27&1.0564&0.0598&24.37&19.57&21.50&1.86&16.73&6.1&r&60&O17\\
18&2025-09-28&1.0616&0.0655&23.70&19.10&21.66&1.71&16.92&8.2&r&70&N54\\
19&2025-09-28&1.0603&0.0641&23.86&19.24&21.60&1.50&16.87&0.9&B/V/R/I&50&O49\\
20&2025-09-29&1.0673&0.0718&23.03&18.65&21.78&1.47&17.11&8.3&r&80&N54\\
21&2025-09-30&1.0733&0.0782&22.41&18.24&21.88&1.28&17.29&7.9&r&100&N54\\
22&2025-10-01&1.0790&0.0845&21.87&17.91&21.94&1.12&17.45&7.5&r&100&N54\\
\hline
\end{tabular}
\label{tab:2}
\raggedright
\footnotesize
\begin{minipage}{\textwidth}
${*}$  Gaoyazi Observatory (no MPC code assigned): 88.577$^{\circ}$ E, 43.522$^{\circ}$ N, 1783 m elevation.\\
$^{\dagger}$ The exposure time for B-band observations was 20 s on 2025 September 24 and 65 s on 2025 September 28.
\end{minipage}
\end{table*}

\section{spin-state and shape}

\subsection{Rotation-period analysis}

To determine the rotation period of 2025~$\mathrm{FA}_{22}$, we selected six consecutive nights of r-band observations from station N54 (23--28 September 2025; lightcurves 15, 16, 18, 20--22 in Table~\ref{tab:2}). These data provided continuous nightly coverage with homogeneous photometry. We derived the synodic period using Fourier series analysis following \citet{pravec2000}, searching for the period that minimizes the $\chi^2$ residuals of the composite lightcurve.

\begin{equation}
M_{i, j}=\sum_{k=1}^{N}\left[B_{k} \sin \frac{2 \pi k(t_{j}-t_{0})}{P}+C_{k} \cos \frac{2 \pi k(t_{j}-t_{0})}{P}\right]+Z_{i}.
\end{equation}

where $M_{i,j}$ are the r-band magnitudes measured at the epoch of $t_j$; $B_k$ and $C_k$ are the Fourier coefficients for $k$-th harmonic; $P$ is the rotation period; $t_0$ is the reference epoch ($\mathrm{JD}_0$); $N$ is the Fourier order; and $Z_i$ are the zero-point offsets accounting for differences between observing nights, CCD chips, and telescope pointings. 

To determine the best Fourier order, we tested orders ranging from 2 to 6 and selected the lowest order for which the root mean square (RMS) of residuals was minimized and the lightcurve morphology was reproduced \citep{pravec2000}. Figure~\ref{fig:2} shows the period-search results, with $\chi^2$ plotted against trial period. The best-fitting synodic rotation period is $P_{\rm synodic} = 13.0795\pm$0.0017~h, and the corresponding Fourier-series fit gives a lightcurve amplitude of $A = 0.59\pm0.04$~mag. The synodic period closely approximates the sidereal period \citep{gehrels1977}, and thus serves as the initial constraint for the period search in the shape inversion (Section~\ref{sec:convex}).

\begin{figure}
\center
	\includegraphics[width=0.9\columnwidth]{synodic_rotation_2025FA221.png}
    \caption{The upper panel shows the $\chi^2$ distribution of trial rotation periods fitted using a sixth-order Fourier series, with the minimum $\chi^2$ corresponding to a synodic rotation period of $P = 13.0795 \pm 0.0017$~h. The lower panel shows the folded lightcurve of 2025 $\mathrm{FA}_{22}$ at an epoch of $\mathrm{JD}_0 = 2460945.0965$, with different markers representing photometric data from different observing nights (see Table ~\ref{tab:2}). The red curve shows the sixth-order Fourier fit with an amplitude of $0.59 \pm 0.04$~mag at $\alpha = 25.5^\circ$.}
    \label{fig:2}
\end{figure}

\subsection{Sidereal period, spin axis and shape}
\label{sec:convex}

Following the standard lightcurve inversion methodology established by \citet{kaasalainen2001a, kaasalainen2001b}, we derived the sidereal period, pole orientation, and convex shape of 2025~$\mathrm{FA}_{22}$ using all lightcurves listed in Table~\ref{tab:2}. The inversion begins with a period search using six trial pole orientations uniformly distributed on the celestial sphere. For each trial period, convex shape models are fitted to the observed lightcurves for all six poles, and the solution with minimum $\chi^2$ is retained. The period values we scanned ranged from 12.0 to 14.0 h, encompassing the synodic period determined from the Fourier analysis (see Fig.~\ref{fig:2}). The result of the period search indicated a primary rotational period at 13.0721 h, which was then used as the initial sidereal period.

The second step is to determine the pole orientation. Our approach involves setting up a grid of pole positions covering the entire celestial sphere with a resolution of $3^\circ \times 3^\circ$. At each pole position, the period and convex shape were optimized to fit the lightcurves. The sidereal period determined in the first step is utilized here as an optimal starting point for the subsequent optimization process. The reference epoch was fixed at $T_0 = 2\,460\,937.0$ JD (18 September 2025, the date of the first observation) with initial rotation phase $\phi_0 = 0^{\circ}$.

\begin{figure}
\center
	\includegraphics[width=0.95\columnwidth]{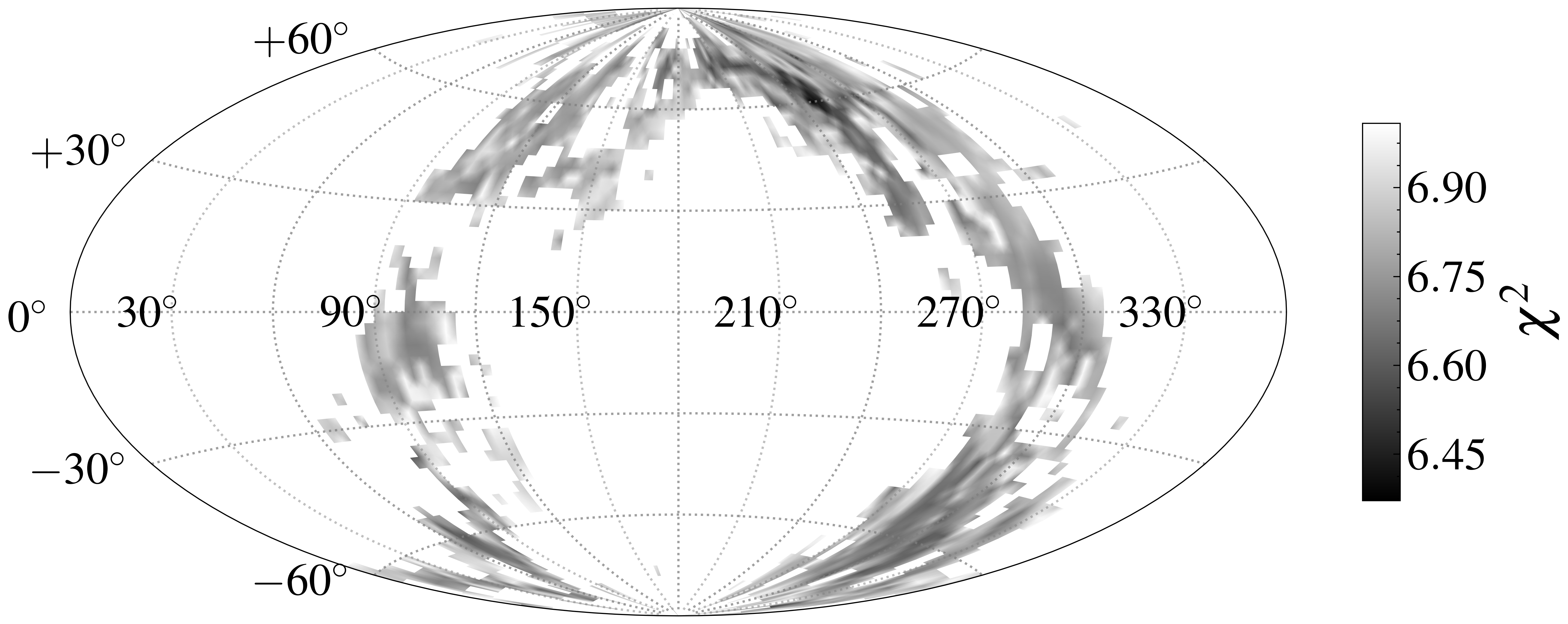}
    \caption{Results of the pole orientation search. $\chi^2$ values are plotted across the celestial sphere in ecliptic coordinates using an Aitoff projection. Darker colours indicate better fits. Any white areas are more than 1.15 times the minimum calculated value. The best pole is found at $(\lambda, \beta) = (246^\circ, 60^\circ)$, with a $1\sigma$ error of radius $9^\circ$.}
    \label{fig:3}
\end{figure}

Following the approach described in \citet{Vok2011} and \citet{Pol2014}, we identified the unique pole solution as having $\chi^2_{\rm min}$ significantly lower than all other solutions, with the threshold defined as $\chi^2_{\rm thr} = (1 + \sqrt{2/\nu})\chi^2_{\rm min}$, where $\nu \approx 2600$ is the number of degrees of freedom. This criterion also defines the $1\sigma$ uncertainty region for the pole orientation. Figure~\ref{fig:3} shows the $\chi^2$ distribution across the celestial sphere.

\begin{figure}
\center
	\includegraphics[width=0.9\columnwidth]{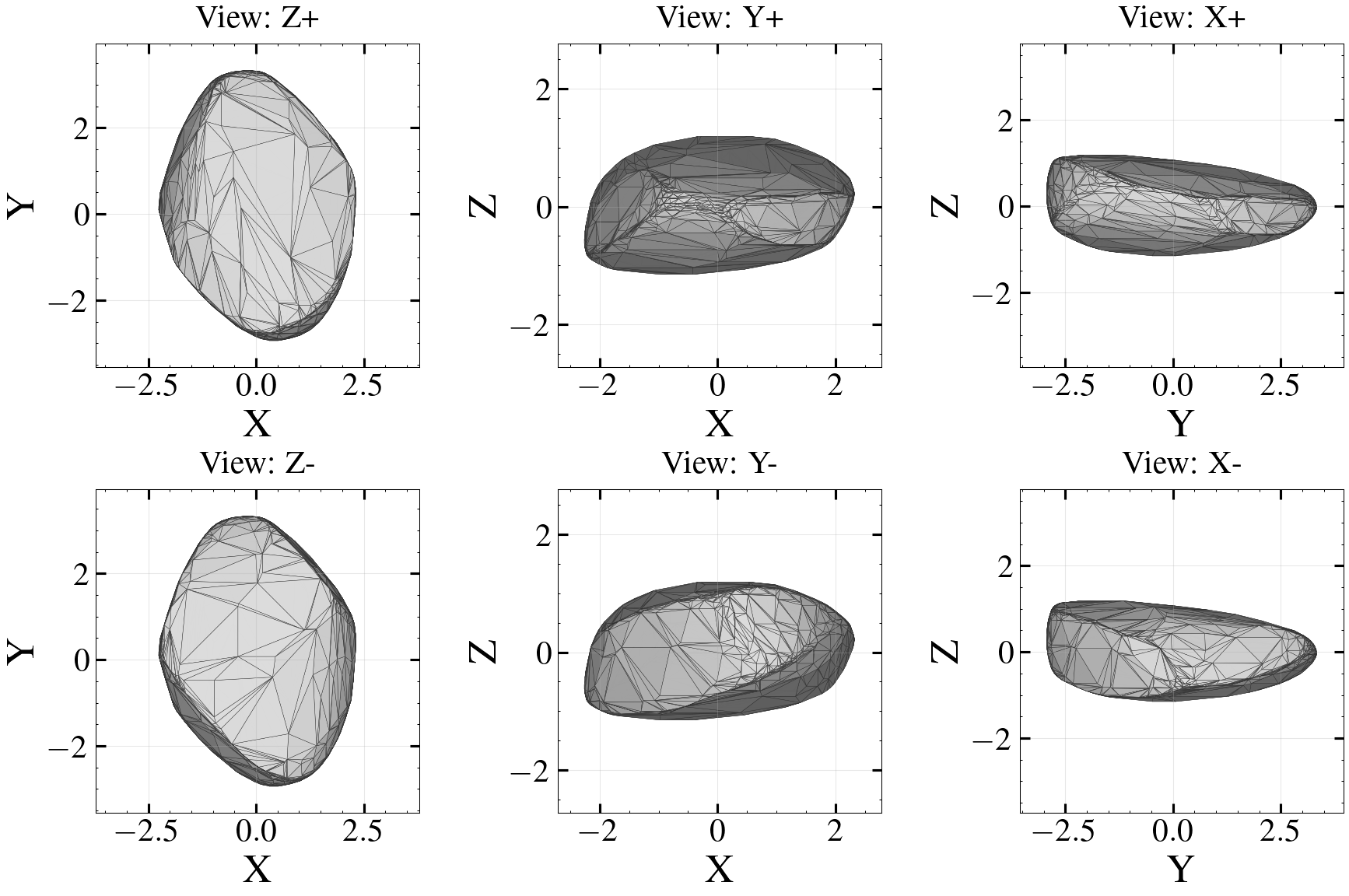}
    \caption{Best-fit convex shape model of 2025 $\mathrm{FA}_{22}$. The model was derived from a pole search using photometric light-curve data under the assumption of a constant rotation period. The sidereal period is $P_{\rm{sid}} = 13.07366$ h and the pole is located at ($\lambda$, $\beta$) = ($246^\circ, 60^\circ$). Top row (left to right): views along the Z, Y and X axes of the body-centric coordinate frame from the positive end of the axis. Bottom row (left to right): views along the Z, Y and X axes from their negative ends. Under the assumption of principal-axis rotation, the Z-axis (spin axis) is aligned with the axis of minimum moment of inertia. The X-axis is arbitrarily oriented such that it lies in the plane of the sky at epoch $T_0$, with no physical correspondence to the principal axes of inertia. All dimensions are relative, as the light-curve inversion method does not constrain the absolute size of the body.}
    \label{fig:4}
\end{figure}

While ecliptic-limited observations typically suffer from a $\lambda \pm 180^{\circ}$ pole ambiguity \citep{Kaasal2006}, our $\chi^2$ map (Figure~\ref{fig:3}) shows a single, well-defined global minimum with no statistically acceptable secondary solution at $\lambda + 180^{\circ}$. The $\chi^2$ value at the mirror position $(\lambda + 180^{\circ}, \beta)$ lies above the global minimum by more than the $1\sigma$ confidence level, so this solution is statistically rejected. The suppression of the mirror solution is due to the broad range of solar phase angles and aspect angles covered by our dataset, which breaks the $\lambda \pm 180^{\circ}$ degeneracy. The best-fit pole solution is at ecliptic coordinates $(\lambda, \beta) = (246^{\circ}, +60^{\circ})$ with a $1\sigma$ uncertainty radius of $9^{\circ}$. At this pole orientation, we refined the sidereal period to $P_{\rm sid} = 13.07366 \pm 0.00076$~h, assuming constant rotation (YORP analysis requires longer temporal baselines). Figure~\ref{fig:4} presents the best-fit convex shape model, which reproduces all observed lightcurves (Appendix~\ref{app:lightcurves}).

The axis ratio of the equivalent ellipsoid is $a{:}b{:}c \sim 2.68{:}1.96{:}1.00$, indicating a elongated shape. The planar features arise from the large light curve amplitudes, while the flattened shape reflects poor constraints along the rotation axis \citep{Durech2020}. Robust shape determination typically requires photometric observations spanning more than $120^{\circ}$ in ecliptic longitude \citep{Durech2020,kwiatk2021}, in order to ensure sufficient variation in viewing geometry. Our observations span only $\sim 100^{\circ}$ in ecliptic longitude, because the first run on 17 September 2025 was excluded owing to its low SNR, which is below this threshold. However, two factors support the robustness of our derived spin parameters and shape model: wide phase angle coverage ($\alpha = 22^{\circ}$--$68^{\circ}$), and substantial viewing geometry variation as the asteroid moved from opposition to elongation. This combination, rarely achieved in single-apparition observations, provides strong constraints that help resolve degeneracies in the shape solution.

\section{Physical characteristic}

\subsection{Colour indices and taxonomy}
\label{sub4.1}
We obtained multi-colour photometry of 2025 $\mathrm{FA}_{22}$ on 24 and 28 September 2025 (IDs 12 and 19). The observations were conducted independently on each night, and the data were reduced following the procedures described in Section 2. On each night, we obtained six sets of observations following the filter sequence B-V-B-V-R-V-R-I-R-I. The colour indices $B-V$, $V-R$, and $R-I$ were computed for each set and then averaged over all six sets. The mean colours of 2025 FA$_{22}$ are $B-V = 0.71 \pm 0.06$ mag, $V-R = 0.38 \pm 0.03$ mag, and $R-I = 0.36 \pm 0.04$ mag on 24 September, and $B-V = 0.71 \pm 0.05$ mag, $V-R = 0.41 \pm 0.03$ mag, and $R-I = 0.41 \pm 0.04$ mag on 28 September.

To determine the taxonomic type of 2025 $\mathrm{FA}_{22}$, we converted the colour indices to the reflection coefficients $R_{\rm{B}}$, $R_{\rm{R}}$ and $R_{\rm{I}}$, normalized to the reflection values in the V band, using the formula given in \citet{DEMC2013}.

\begin{equation}
    R_{f}=10^{-0.4\left[\left(M_{f}-M_{\rm{V}}\right)-\left(M_{f, \odot}-M_{\rm{V}, \odot}\right)\right]}.
\end{equation}

where $M_{f}$ and $M_{f, \odot}$ are the magnitudes of the object and sun in a certain filter $f$, respectively, at the central wavelength of the filter. The equation is normalized to unity at the central wavelength of filter V using $M_{\rm{V}}$ and $M_{\rm{V}, \odot}$: the V magnitudes of the object and sun, respectively. For the solar magnitudes, we adopt the values provided by \citet{willmer2018}: $M_{\rm{B},\odot}=-26.13$, $M_{\rm{V},\odot}=-26.76$ mag, $M_{\rm{R},\odot}=-27.15$ mag, and $M_{\rm{I},\odot}=-27.47$ mag. 

Next, We classified 2025~FA$_{22}$ using the Bus-DeMeo taxonomic system \citep{DeMeo2009}. Our multi-colour B/V/R/I photometry provides reflectance measurements at discrete wavelengths, which we linearly interpolated onto a uniform grid (0.45--0.80 $\mu$m, 0.05 $\mu$m intervals) to match the sampling of the Bus-DeMeo standard spectra. Since broadband photometry lacks the spectral resolution required to distinguish taxonomic subtypes, we compared our interpolated spectra against the mean spectra of the major complexes rather than individual subtypes, and calculated the Euclidean distance as a measure of similarity. This analysis was performed independently for observations obtained on 24 and 28 September 2025. 

\begin{figure}
\center
    \includegraphics[width=0.95\columnwidth]{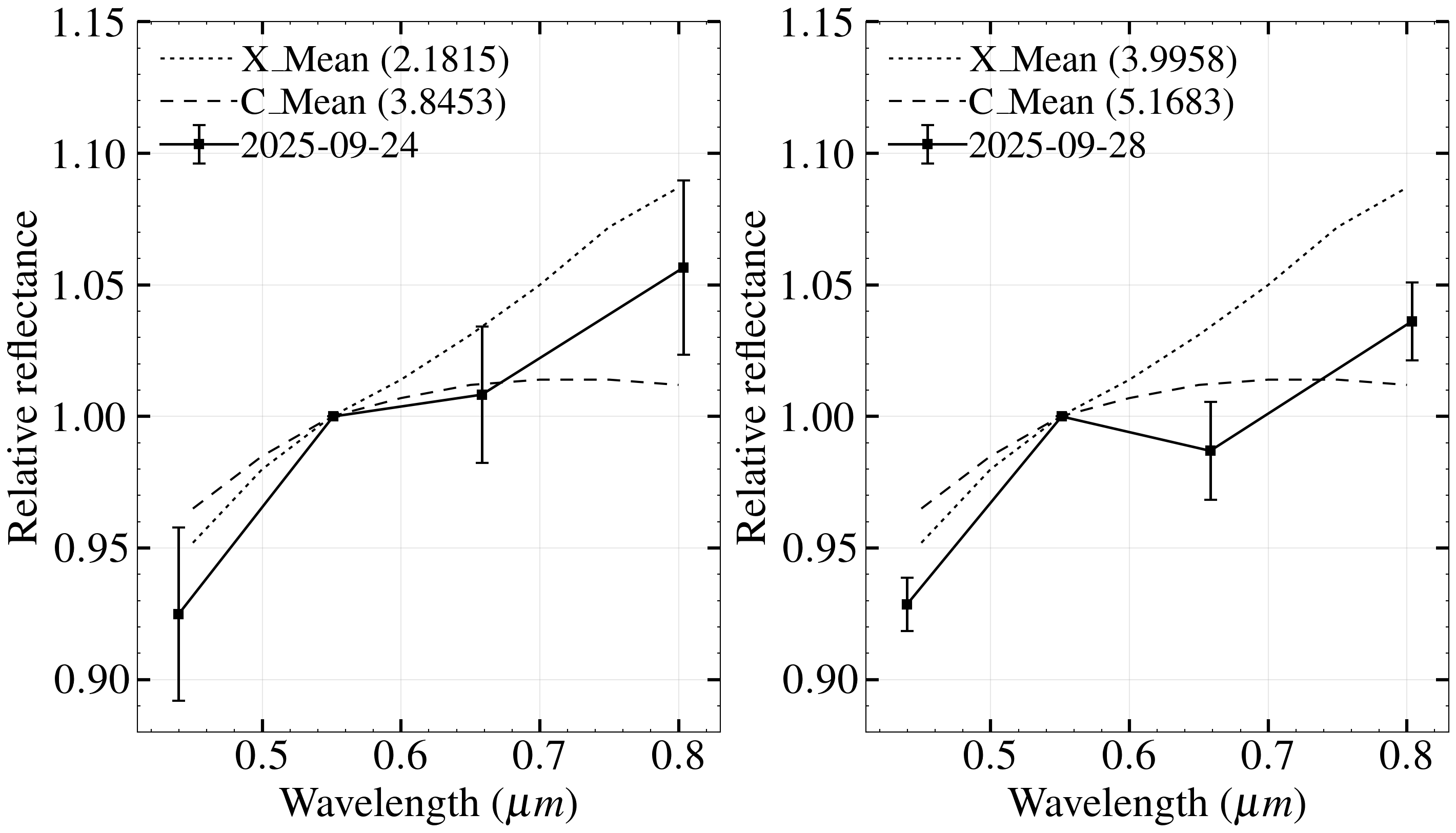}
    \caption{Taxonomic classification of 2025~FA$_{22}$ based on minimum-distance matching to Bus-DeMeo standard spectra \citep{DeMeo2009}. Left and right panels show observations from 24 and 28 September 2025, respectively. Black squares with error bars represent B,V,R, and I photometry normalised to unity at V-band. Dashed and dotted lines show the mean spectra of the X and C complexes, respectively, with Euclidean distances given in parentheses. The X complex provides a closer match on both nights.}
    \label{fig:6}
\end{figure}

Figure~\ref{fig:6} shows the comparison between our observed photometry and the mean spectra of the X and C complexes in the Bus-DeMeo taxonomy \citep{DeMeo2009}, with Euclidean distances given in parentheses. The X complex is associated with the smallest distances on both nights, and is therefore adopted as the most probable taxonomic classification. This is broadly consistent with the visible spectrum obtained by Camacho et al. on 21 September 2025\footnote{\url{https://iawn.net/obscamp/2025FA22/2025FA22\_gallery.shtml}}, which was reported in the IAWN campaign gallery and classified as Xk type. This classification carries implications for hazard assessment, as X-complex asteroids span a wide range of compositions: they may be carbonaceous bodies ($\rho \sim 2$--$3~\mathrm{g~cm^{-3}}$) or iron-rich objects ($\rho \sim 7$--$8~\mathrm{g~cm^{-3}}$) \citep{DeMeo2022}. The latter would make 2025~FA$_{22}$ a particularly hazardous PHA, since iron-rich bodies are highly resistant to atmospheric disruption and deliver their kinetic energy directly to the surface \citep{Melosh2005}.

Resolving this ambiguity requires near-infrared spectroscopy ($\lambda > 1~\mu$m) to search for silicate absorption features characteristic of carbonaceous material, or confirm the featureless spectrum expected for metal-rich compositions \citep{shepard2015}. We recommend such observations during the 2036 apparition (Section~\ref{sec:Discussionc}) to constrain the composition and refine the impact hazard posed by this object.

\subsection{Magnitude-phase curve}

The absolute magnitude $H$ of a small body is defined as its reduced magnitude at zero phase angle. Since this geometry (Sun--asteroid--Earth alignment) cannot be directly observed, $H$ is always determined by fitting a photometric model to the observed magnitude–phase angle relation and evaluating it at $\alpha = 0^{\circ}$ \citep{Bowell1989,Muinonen2010}. Its accuracy therefore depends strongly on the phase-angle coverage of the observations, which is often limited for NEAs because small phase angles are rarely sampled, the viewing geometry changes rapidly during close approaches, and rotational brightness variations may distort the phase curve.

Rotational brightness variations can distort the magnitude--phase angle relation and must therefore be removed before constructing the phase function of 2025~FA$_{22}$. For each observing night, we fitted the lightcurve with a Fourier series (typically 4th--6th order). The rotationally averaged magnitude was then taken as the mean of the fitted function over one full rotation cycle, thereby removing shape-induced brightness variations. The $1\sigma$ standard deviation of the fitted function over the same cycle was adopted as the uncertainty on the averaged magnitude. Only nights with sufficient rotational phase coverage were retained, since a reliable mean cannot be obtained from an incomplete lightcurve. Observations obtained before 20 September 2025 were also excluded, as the viewing geometry changed substantially during that period (see Figure~\ref{fig:1}).

From the remaining data, we selected 11 lightcurves (IDs 8--11, 13, 15, 16, 18, 20--22; Table~\ref{tab:2}), each from a separate observing night. All were obtained in the $r$ band and calibrated against solar-type stars from the SDSS DR12 catalogue \citep{sdss2015}. The solar phase angle for each night was computed at the mid-time of the lightcurve. The resulting rotationally averaged magnitudes and their uncertainties are listed in Table~\ref{tab:phase}, and each serves as a single data point in the magnitude--phase angle diagram.

\begin{table}
\centering
\caption{Phase curve data for 2025~FA$_{22}$. The Julian date corresponds to the mid-time of each lightcurve. The uncertainty on each averaged magnitude is taken as the $1\sigma$ standard deviation of the fitted lightcurve function over the same rotation cycle.}
\label{tab:phase}
\begin{tabular}{ccccc}
\hline
IDs & JD (mid) & $\alpha$ (deg) & $r(1,1,\alpha)$ (mag) & $\sigma$ (mag) \\
\hline
8  & 2460940.31435  & 33.11 & 22.75 & 0.20 \\
9  & 2460941.34642  & 30.46 & 22.51 & 0.25 \\
10 & 2460942.33348  & 28.71 & 22.56 & 0.22 \\
11 & 2460943.29722  & 27.38 & 22.44 & 0.19 \\
13 & 2460944.28544  & 26.26 & 22.45 & 0.18 \\
15 & 2460945.26452  & 25.31 & 22.33 & 0.22 \\
16 & 2460946.28020  & 24.44 & 22.42 & 0.23 \\
18 & 2460947.26452  & 23.70 & 22.29 & 0.26 \\
20 & 2460948.24980  & 23.03 & 22.32 & 0.18 \\
21 & 2460949.26261  & 22.41 & 22.29 & 0.17 \\
22 & 2460950.24119  & 21.87 & 22.32 & 0.18 \\
\hline
\end{tabular}
\end{table}

After scaling the magnitudes to unit heliocentric and geocentric distances, we fitted the phase curve with the three-parameter $H$--$G_1$--$G_2$ model \citep{Muinonen2010}. This nonlinear model is particularly suited for determining the phase curves of near-Earth objects (NEOs), providing reliable parameter estimates even when the data contain gaps in phase-angle coverage or exhibit large scatter in the individual data points\citep{Penttila2016}. In reduced-magnitude space, the model is given by:

\begin{equation}
\begin{aligned}
V(1,1,\alpha) = H - 2.5 \log_{10} \bigl[ \,
& G_1 \Phi_1(\alpha) + G_2 \Phi_2(\alpha) \\
& + (1 - G_1 - G_2)\Phi_3(\alpha) \bigr],
\end{aligned}
\end{equation}

where $\alpha$ is the solar phase angle, $V(1,1,\alpha)$ is the reduced magnitude at phase angle $\alpha$, $H$ is the absolute magnitude, and $G_1$ and $G_2$ are the phase-slope parameters. The basis functions $\Phi_1$, $\Phi_2$, and $\Phi_3$ are described in detail by \citet{Muinonen2010}. We performed the fit using the online tool provided by \citet{Penttila2016},\footnote{Available at \url{https://psr.it.helsinki.fi/HG1G2/}.} which implements the three-parameter $H$--$G_1$--$G_2$ system.

\begin{figure}
\center
	\includegraphics[width=0.95\columnwidth]{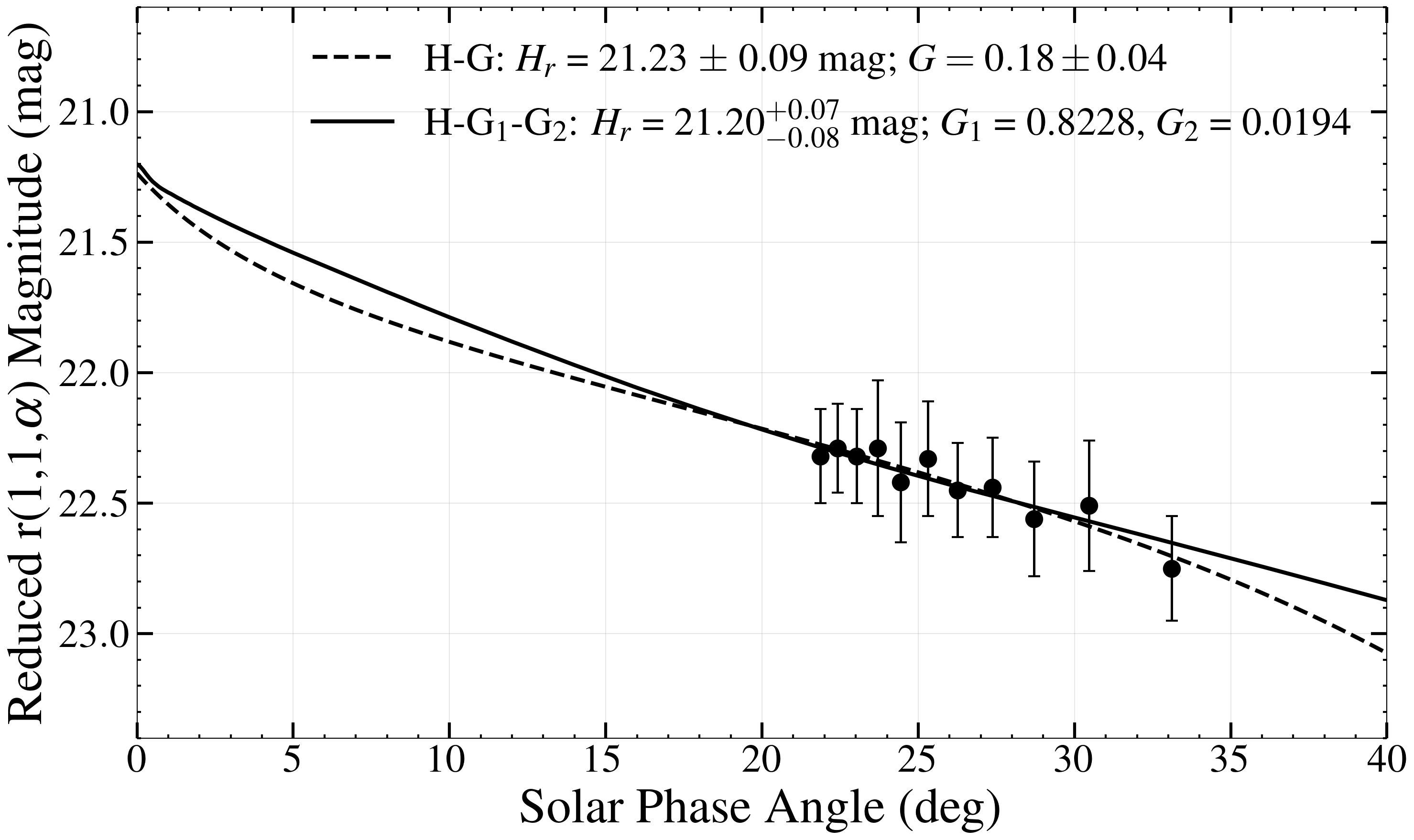}
    \caption{Magnitude-phase curve of 2025 FA$_{22}$. The solid line shows the $H$--$G_1$--$G_2$ fit, giving $H_{\mathrm{r}}=21.20^{+0.07}_{-0.08}$~mag, $G_1=0.8228$, and $G_2=0.0194$. The dashed line shows the $H$--$G$ fit, with $H_{\mathrm r}=21.23\pm0.09$~mag and $G=0.18\pm0.04$.}
    \label{fig:7}
\end{figure}

The nonlinear constrained least-squares fit was applied to 11 data points spanning a phase angle range of $21.87^{\circ}$--$33.11^{\circ}$ (Table~\ref{tab:phase}). The Bayesian Information Criterion (BIC) favours the one-parameter phase-function model based on the Tholen C-type template. In this solution, the phase-curve shape is fixed to that of C-type asteroids in the Tholen taxonomy \citep{Tholen1984}, broadly encompassing the Bus-DeMeo C-complex subtypes B, Cb, Cg, Cgh, and Ch, with $G_1 = 0.8228$ and $G_2 = 0.0194$. This gives $H_{\mathrm{r}} = 21.20^{+0.07}_{-0.08}$~mag and a phase coefficient of 0.0328~mag$\cdot$deg$^{-1}$ (Fig.~\ref{fig:7}). Allowing all three parameters to vary freely leads to the boundary solution $G_2 = 0$, indicating that the present phase-angle coverage does not independently constrain $G_2$.

For comparison, we also fitted the data with the classical two-parameter $H$--$G$ phase function \citep{Bowell1989}, obtaining $H_{\mathrm{r}} = 21.23 \pm 0.09$~mag and $G = 0.18 \pm 0.04$ (Fig.~\ref{fig:7}). We converted $H_{\mathrm{r}}$ to the standard $V$ band using the color transformation of \citet{Jester2005}, $H_{V} = H_{\mathrm{r}} + 0.42(B-V) - 0.11$, where $B-V$ is adopted from Section~\ref{sub4.1}. This gives $H_{\mathrm{V}} = 21.39^{+0.07}_{-0.08}$~mag from the $H$--$G_1$--$G_2$ fit and $H_{\mathrm{V}} = 21.42 \pm 0.10$~mag from the $H$--$G$ fit.

\section{Discussion}
\label{sec:Discussionc}

With the derived absolute magnitude of $H_{\mathrm{V}} = 21.39^{+0.07}_{-0.08}$~mag, we estimate the effective diameter of 2025~FA$_{22}$ using 
the standard equation $D_{\mathrm{eff}}=1329\times 10^{(-H_{\mathrm{V}}/5)} \times p_v^{-1/2}$ \citep{harris1997}. Based on its multicolour photometry 
(Section~\ref{sub4.1}), 2025~FA$_{22}$ is classified as X-complex. Within the taxonomy of \citet{mahlke2022}, the X-complex spans a wide albedo range, from high-albedo E-type through moderate-albedo M-type to low-albedo P-type. We therefore consider two representative albedo assumptions bracketing this range \citep{mahlke2022}. Adopting the mean M-type albedo to represent a moderate-albedo case, $p_v = 0.15^{+0.05}_{-0.04}$, gives $D_{\mathrm{eff}} = 181^{+31}_{-25}$ m. Using the mean P-type albedo to represent a low-albedo case, $p_v = 0.05^{+0.02}_{-0.02}$, gives $D_{\mathrm{eff}} = 313^{+64}_{-64}$ m. The Goldstone Solar System Radar had scheduled observations of 2025 FA$_{22}$, but the observing window was lost due to an unexpected technical failure\footnote{\url{https://echo.jpl.nasa.gov/asteroids/2025fa22.2025.goldstone.planning.html}}. The “China Compound Eye” Radar acquired delay-Doppler images revealing a distinct bifurcated shape, indicating a likely contact binary morphology. Radar observations yield approximate dimensions of $100 \times 320$~m (Zehua Dong, private communication), corresponding to an effective diameter of $\sim$186~m. This is in good agreement with our moderate-albedo estimate ($D_{\mathrm{eff}} = 181^{+31}_{-25}$~m, $p_v = 0.15$), supporting a moderate geometric albedo for 2025~FA$_{22}$ and consistent with its X-complex classification.

By combining the radar-derived diameter $D_{\mathrm{eff}} = 186$~m with our determined absolute magnitude $H_{\mathrm{V}} = 21.39^{+0.07}_{-0.08}$~mag, we derive a geometric albedo of $p_v\approx0.14$ for 2025~FA$_{22}$. \citet{Rondon2022} and \citet{Belskaya2000} reported a significant correlation between phase coefficients and albedo, providing a fitted relationship between the two parameters. Using the phase coefficient 0.0328~mag$\cdot$deg$^{-1}$ derived from our $H$--$G_1$--$G_2$ phase-curve fit, we estimate a geometric albedo of $p_v\approx0.15$ for 2025~FA$_{22}$. This result is consistent with both the geometric albedo independently derived from the combination of radar measurements and absolute magnitude and the mean albedo of X-complex asteroids indicated by our multicolor photometry, further supporting the robustness of our analysis.

The bifurcated shape revealed by the radar delay-Doppler images, together with the physical properties derived above, motivates a closer examination of the formation pathways that could produce such a contact-binary configuration. Contact binaries can form through several mechanisms: (1) YORP-driven spin-up and fission of monolithic bodies, or gradual reshaping of heterogeneous asteroids \citep{Jacobson2011,Scheeres2018}; (2) BYORP-driven orbital decay and merger in binary systems \citep{Cuk2005}; (3) collisional re-accumulation following impacts \citep{Cannon2025,michel2013}; (4) low-velocity mergers in the outer Solar System \citep{jutzi2015,campo2020}; (5) secondary formation in multi-body systems \citep{Levison2024}. For 2025 FA$_{22}$, mechanisms (4) and (5) are inapplicable given its inner Solar System environment and lack of evidence for multi-body association. Collisional re-accumulation (mechanism 3) is unlikely for a $\sim$160-m NEA because gravitational re-accumulation into contact binaries requires low-velocity environments ($v < 1$ m/s) found in post-collision debris fields, not the high-velocity ($>$ several km/s) collisional regime typical of NEAs \citep{campo2020}. The BYORP-driven mechanism is also unlikely given that timescales for a 166-m object are comparable to or exceed NEA dynamical ages ($\sim$few Myr), insufficient for orbital decay and merger \citep{Cuk2005}.

The YORP-driven mechanism provides the most straightforward explanation for the formation of 2025~FA$_{22}$. The contact binary morphology could have formed during a previous YORP cycle when the object rotated faster, potentially approaching the rotational disruption limit. The current slow rotation period ($P \sim 13$ h) may result from subsequent YORP deceleration or gravitational perturbations during close planetary encounters. Given that contact binaries have strongly irregular shapes, substantial thermal torques are expected, particularly for sub-kilometre NEAs. Using the JPL Horizons orbit solution 68 (JPL\#68; heliocentric IAU76/J2000 ecliptic osculating elements, accessed on 2026-01-21) to define the orbital pole, together with the spin-axis orientation derived in this work, we compute the obliquity (the angle between the spin axis and the orbital angular-momentum vector) as $\varepsilon = 37^{\circ}$. This differs from the obliquities of contact binaries such as 1999~JV$_6$ ($\varepsilon = 174^{\circ}$; \citealt{Rozek2019}) and (25143)~Itokawa ($\varepsilon = 178^{\circ}$; \citealt{scheeres2008}), which lie close to the $\varepsilon = 180^{\circ}$ attractor, and may indicate a less evolved state within the YORP cycle. This evolutionary stage makes 2025~FA$_{22}$ a particularly valuable target for investigating the YORP cycle and the formation pathways of contact binaries.

Observations of 2025 FA$_{22}$ to date cover only a single opposition. The shape model derived from this photometry alone cannot constrain formation mechanisms or detect YORP-induced changes in the spin state. However, 2025 FA$_{22}$ adds to the growing sample of contact binaries with reliable physical parameters and preliminary shape model, enabling comparative studies across rotational states, spectral types, and size ranges to address whether such objects form primarily through YORP-driven evolution or through alternative pathways. 

For a $\sim$160~m NEA like 2025~FA$_{22}$, YORP signatures may become detectable at the next opposition. The 2036 close approach will bring the asteroid to 0.13~au from Earth\footnote{\url{https://ssd.jpl.nasa.gov/tools/sbdb\_lookup.html\#/?sstr=2025\%20FA22\&view=OPC}}. While too distant for high-quality radar observations, this provides an opportunity for optical follow-up to refine the spin state. Lightcurves obtained 11~yr after our 2025 observations would enable detection of YORP-induced period changes. The 2036 viewing geometry ($\alpha = 131^{\circ}$, $\lambda_{\mathrm{O}} = 110^{\circ}$, $\beta_{\mathrm{O}} = -22^{\circ}$) will differ substantially from our current dataset (Table~\ref{tab:2}), enabling better pole constraints. However, the asteroid will reach only $V=20$~mag at maximum brightness, requiring telescopes with apertures $\gtrsim 1$ m. The 4.2-m dedicated astrometric telescope under construction in China (expected completion in 2027) will be capable of such observations\footnote{\url{https://pmo.cas.cn/xwdt2019/kyjz2019/202506/t20250624\_7873950.html}}. 

\section{Conclusions}

We conducted photometric observations of 2025 FA$_{22}$, the 7th IAWN Campaign target, during its 2025 apparition. A photometric reduction pipeline for asteroid photometry has been developed and is applicable to various small-aperture telescopes. The sidereal rotation period of 2025 FA$_{22}$ is $P_{\mathrm{sid}}=13.07366\pm0.00076 $ h, with a spin axis orientation of $(\lambda, \beta) = (246\pm 9^\circ, 60\pm 9^\circ)$. Convex shape inversion shows an elongated ellipsoidal shape with axis ratios of $a{:}b{:}c \sim 2.68{:}1.96{:}1.00$, in good agreement with radar observations. We also derived magnitude--phase parameters $H_{\mathrm{r}} = 21.20^{+0.07}_{-0.08}$~mag ($G_1 = 0.8228$, $G_2 = 0.0194$), which converts to $H_\mathrm{V} = 21.39^{+0.07}_{-0.08}$~mag after applying the $r$-to-$V$ colour transformation. Colour indices measured on 24 September 2025 are $B-V = 0.71 \pm 0.06$~mag, $V-R = 0.38 \pm 0.03$~mag, and $R-I = 0.36 \pm 0.04$~mag; on 28 September 2025 we obtain $B-V = 0.71 \pm 0.05$~mag, $V-R = 0.41 \pm 0.03$~mag, and $R-I = 0.41 \pm 0.04$~mag. The colours measured on both nights are mutually consistent and indicate membership in the X-complex. Adopting a representative X-complex geometric albedo of $p_v = 0.15^{+0.05}_{-0.04}$, we estimate an effective diameter of $D_{\mathrm{eff}} = 181^{+31}_{-25}$~m, consistent with radar-derived size estimates.

The shape model derived from photometry alone is insufficient to constrain the evolutionary pathway of 2025~FA$_{22}$. However, this work expands the sample of contact binaries with well-determined physical parameters and shape models, which is important for assessing the role of YORP-driven evolution in the formation and evolution of near-Earth contact-binary systems. With an increasing number of morphologically similar objects characterised in a homogeneous manner, comparative studies can test whether contact binaries are more consistent with YORP-driven reshaping/fission followed by reconfiguration, or with the evolution and merger of binary systems (mechanism 2). Continued photometric and radar monitoring of 2025~FA$_{22}$ during future apparitions may enable the detection of YORP-induced spin-rate changes and provide improved constraints on its thermal properties and formation history.

\section*{Acknowledgements}

This work is supported by the National Natural Science Foundation of China (grant No. 62227901), the Quantum Science and Technology National Science and Technology Major Project (grant No. 2024ZD0300304), the National Key R\&D Program of China (grant No. 2025YFF0511202), the National Key R\&D Program of China (grant No. 2023YFA1608100), the Science and Technology Project of Qinghai Province (grant No. 2025-ZJ-T01), the Egyptian Science, Technology \& Innovation Funding Authority (STDF, grant No. 48102), the key technology research on the LUV CCD imaging spectrometer (grant No. 12273118), the Jiangsu Province Dual-Creation Doctor Program (grant No. JSSCBS0318) and the Minor Planet Foundation. 

We thank Zehua Dong for providing the radar observations of 2025~FA$_{22}$ and for valuable discussions. This research made use of the NASA/JPL HORIZONS ephemeris system. The $H$--$G_1$--$G_2$ fits were performed using the Online Calculator for Photometric phase-curves (OCP) provided by \citet{Penttila2016}.

%%%%%%%%%%%%%%%%%%%%%%%%%%%%%%%%%%%%%%%%%%%%%%%%%%
\section*{Data Availability}

The photometric lightcurves analysed in this study, together with the derived shape-model files and the {\sc PYTHON~3} data-reduction and analysis scripts, are available from the corresponding author upon reasonable request.

\bibliographystyle{mnras}
\bibliography{example} % if your bibtex file is called example.bib

\appendix
\section{Lightcurve fits from shape inversion}
\begin{figure*}
	\includegraphics[width=0.9\textwidth]{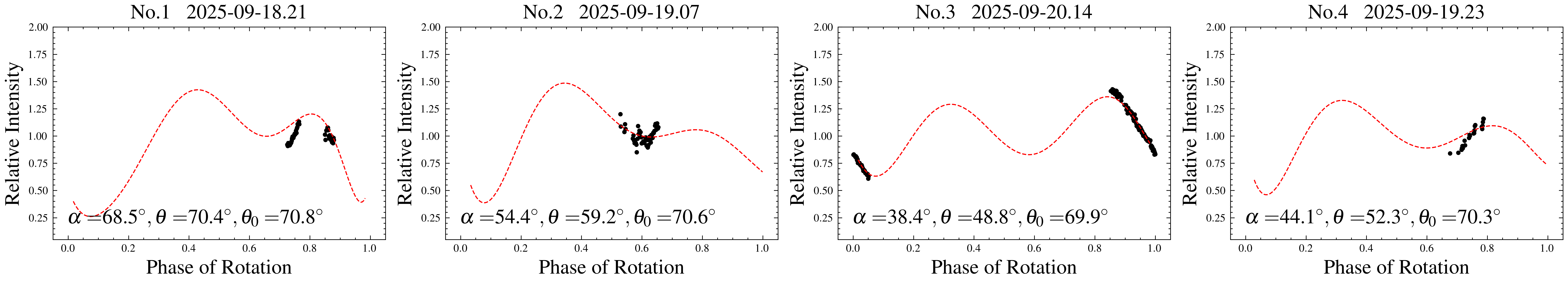}\\
    \includegraphics[width=0.9\textwidth]{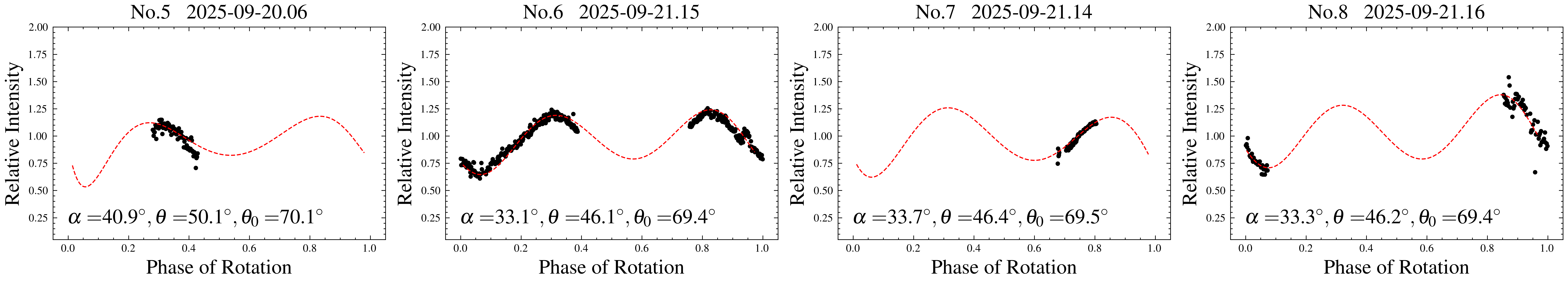}\\
    \includegraphics[width=0.9\textwidth]{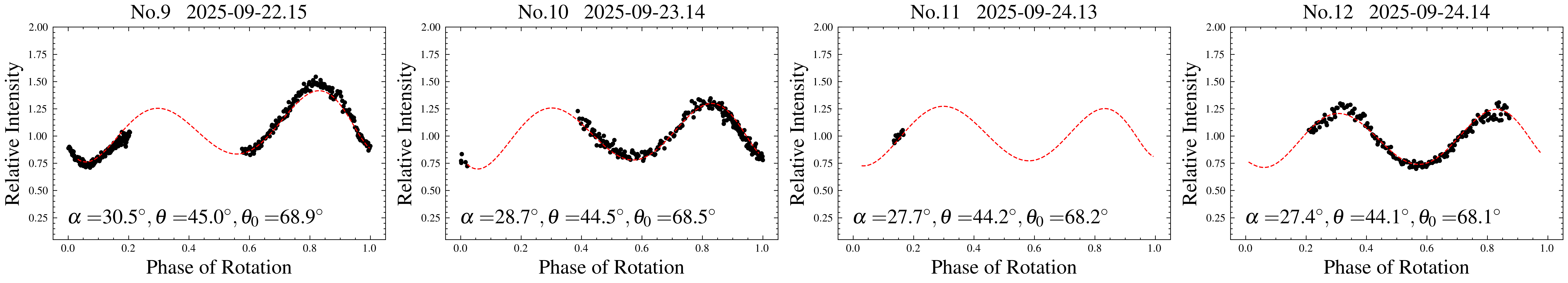}\\
    \includegraphics[width=0.9\textwidth]{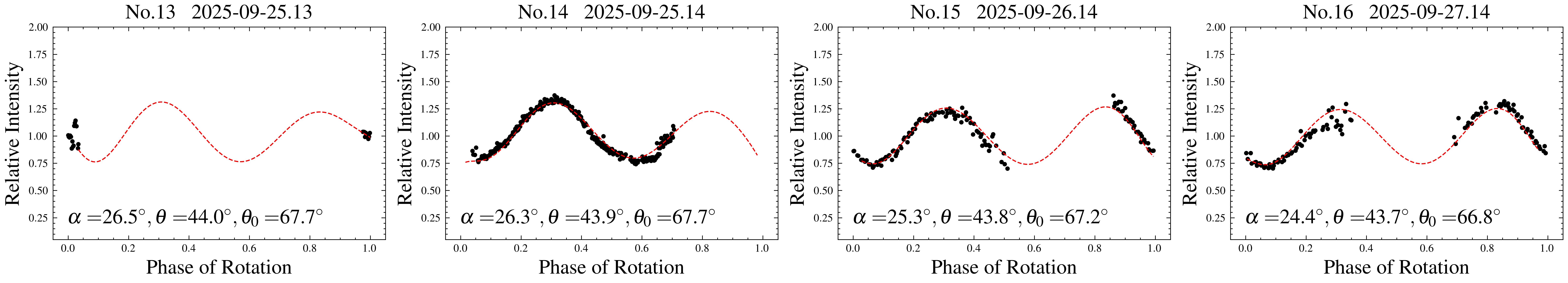}\\
    \includegraphics[width=0.9\textwidth]{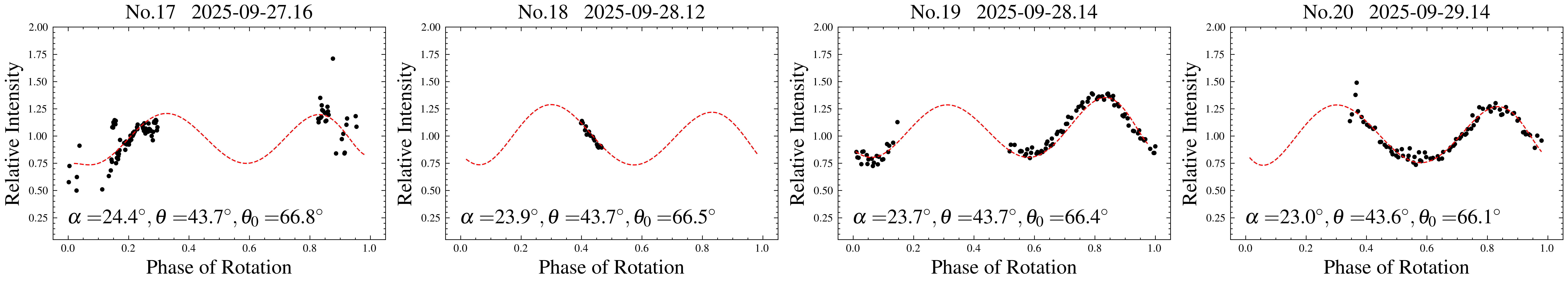} \\
    \includegraphics[width=0.45\textwidth]{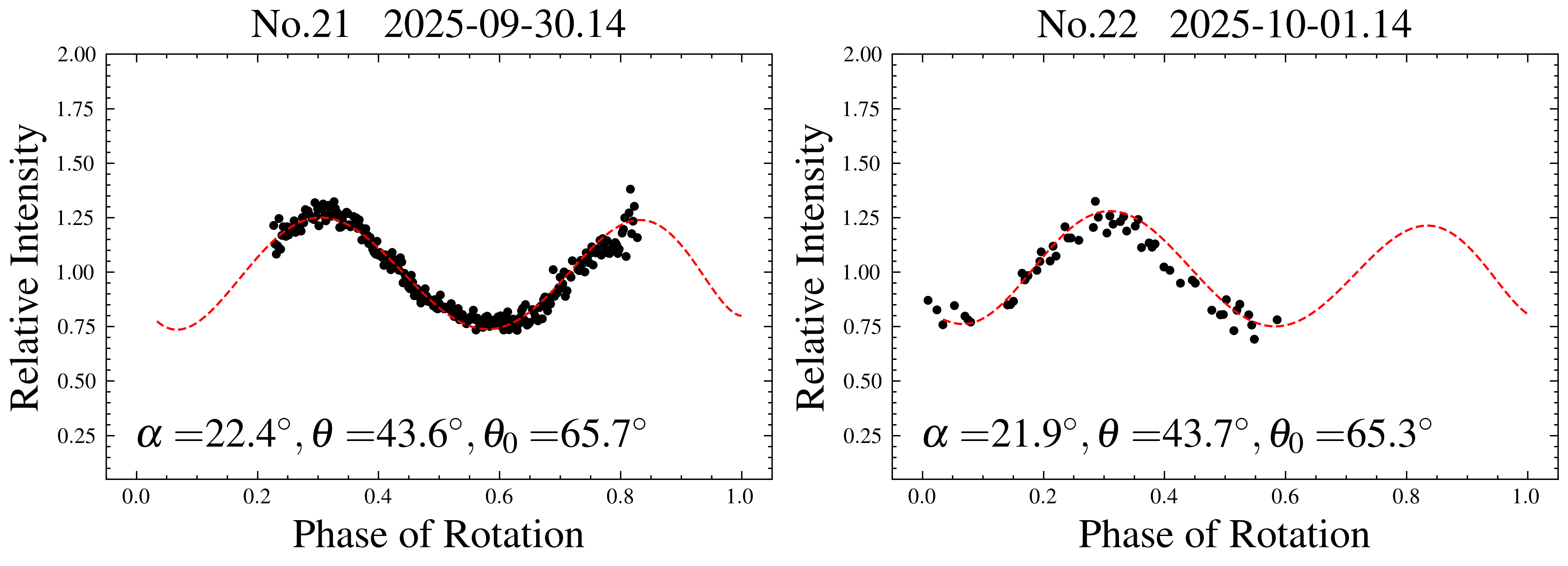}
    \caption{All lightcurves of 2025 $\mathrm{FA}_{22}$ (black dots) alongside the synthetic lightcurves (red dashed curves) produced by the convex shape with spin parameters ($P$, $\lambda$, $\beta$ )=($13.073661$ h, $246^\circ, 60^\circ$). The viewing and illumination geometry is given by the aspect angle $\theta$, the solar aspect angle $\theta_0$, and the solar phase angle $\alpha$.}
    \label{app:lightcurves}
\end{figure*}

The fits to all light curves used in the shape inversion are shown in Figure~\ref{app:lightcurves}. 

%\begin{figure}
%\center
%	\includegraphics[width=0.95\columnwidth]{radar_FA22.png}
%   \caption{Delay-Doppler image of 2025 $\mathrm{FA}_{22}$ from 2025 September 17 observations with the “China Compound Eye” Radar (Zehua Dong, %personal communication). The image shows Doppler shift (vertical axis, increasing left to right) versus time delay (horizontal axis, increasing top %to bottom), with the two-lobed structure consistent with a contact binary.}
%    \label{app:radar}
%\end{figure}

%%%%%%%%%%%%%%%%%%%%%%%%%%%%%%%%%%%%%%%%%%%%%%%%%%

% Don't change these lines
\bsp	% typesetting comment
\label{lastpage}
\end{document}